\documentclass[aps,nolongbibliography, superscriptaddress, twocolumn]{revtex4-2}
\usepackage[UKenglish]{babel}
\usepackage{latexsym}

\usepackage[pdftex, colorlinks=true, linkcolor=red, citecolor=blue, urlcolor=magenta, pdfstartview=FitH]{hyperref}
\usepackage{graphicx}	
\usepackage{amssymb}
\usepackage{amsmath}
\usepackage{xcolor}
\usepackage[normalem]{ulem}

\pdfoutput=1

\usepackage{bm}
\usepackage{verbatim}
\usepackage{amsmath}
\usepackage{amssymb}
\usepackage[utf8]{inputenc}

\usepackage{nicefrac}
\usepackage{braket}
\usepackage{pdfcomment}
\usepackage{pbox}
\usepackage{makecell}
\usepackage{hhline}
\usepackage{multirow}
\usepackage{tabularx}
\usepackage{soul}
\usepackage{float}
\usepackage{microtype}
\usepackage[dvipsnames]{xcolor}

\begin{document}

\title{Enhancement of the WS$_2$ A$_{1\text{g}}$ Raman Mode in MoS$_2$/WS$_2$ Heterostructures}
%
%
\author{Annika Bergmann-Iwe}
\affiliation{Institute of Physics, Rostock University, Germany}
\author{Tomasz Wo\'{z}niak}
\affiliation{Institute of Theoretical Physics, University of Warsaw, Poland}
\affiliation{Institute of Theoretical Physics, Wrocław University of Science and Technology, Poland}
\author{Mustafa Hemaid}
\affiliation{Institute of Physics, Rostock University, Germany}
\author{Ois\'{i}n Garrity}
\affiliation{Department of Physics, Freie Universität Berlin, Germany}
\author{Patryk Kusch}
\affiliation{Department of Physics, Freie Universität Berlin, Germany}

\author{Rico Schwartz}
\affiliation{Institute of Physics, Rostock University, Germany}
\author{Ziyang Gan}
\affiliation{Institute of Physical Chemistry, Friedrich Schiller University Jena, Germany}
\author{Antony George}
\affiliation{Institute of Physical Chemistry, Friedrich Schiller University Jena, Germany}
\author{Ludger Wirtz}
\affiliation{Department of Physics and Materials                     Science, University of Luxembourg, Luxembourg}
\author{Stephanie Reich}
\affiliation{Department of Physics, Freie Universität Berlin, Germany}
\author{Andrey Turchanin}
\affiliation{Institute of Physical Chemistry, Friedrich Schiller University Jena, Germany}
\author{Tobias Korn}
\email{tobias.korn@uni-rostock.de}
\affiliation{Institute of Physics, Rostock University, Germany}

\begin{abstract}
When combined into van der Waals heterostructures, transition metal dichalcogenide monolayers enable the exploration of novel physics beyond their unique individual properties. However, for interesting phenomena such as interlayer charge transfer and interlayer excitons to occur, precise control of the interface and ensuring high-quality interlayer contact is crucial. Here, we investigate bilayer heterostructures fabricated by combining chemical-vapor-deposition-grown MoS$_2$ and exfoliated WS$_2$ monolayers, allowing us to form several heterostructures with various twist angles within one preparation step. In case of sufficiently good interfacial contact – evaluated by photoluminescence quenching - we observe a twist-angle-dependent enhancement of the WS$_2$ A$_{1g}$ Raman mode. In contrast, other WS$_2$ and MoS$_2$ Raman modes (in particular, the MoS$_2$ A$_{1g}$ mode) do not show a clear enhancement under the same experimental conditions. We present a systematic study of this mode-selective effect using nonresonant Raman measurements that are complemented with ab-initio calculations of Raman spectra. We find that the selective enhancement of the WS$_2$ A$_{1g}$ mode exhibits a strong dependence on interlayer distance. We show that this selectivity is related to the A$_{1g}$ eigenvectors in the heterolayer: the eigenvectors are predominantly localized on one of the two layers; yet, the intensity of the MoS$_2$ mode is attenuated because the WS$_2$ layer is vibrating (albeit with much lower amplitude) out of phase, while the WS$_2$ mode is amplified because the atoms on the MoS$_2$ layer are vibrating in phase. To separate this eigenmode effect from resonant Raman enhancement, our study is extended with near-resonant Raman measurements. 

\end{abstract}

\maketitle
\section{Introduction}

Since the discovery of graphene~\cite{Novoselov2004_Science}, two-dimensional (2D) materials have garnered a lot of scientific interest. Besides graphene, transition metal dichalcogenide (TMDC) monolayers, such as MoS$_2$ and WS$_2$ are among the most-investigated 2D materials due to their direct bandgap ~\cite{Mak2010_PhysRevLett, Splendiani2010_NanoLett, Zhao2012_ACS_Nano, Tonndorf2013_OptExpr} and tightly bound excitons~\cite{Molina-Sanchez2015Dec,Chernikov2014_PhysRevLett}. In addition to the unique properties of individual 2D materials, new emergent properties are present when they are combined into van der Waals (vdW) heterostructures~\cite{Geim2013_Nature}. Material combination and stacking sequence alter the heterostructure's electronic properties and are therefore parameters for adapting the heterostructure to a desired application. The relative crystallographic orientation of adjacent layers offers an additional new degree of freedom, which is not accessible in epitaxially grown heterostructures. For example, slightly twisted (so-called magic angle) homobilayers of graphene show exciting transport characteristics including superconductivity~\cite{Cao2018_Nature}.  
In TMDC heterobilayers, inherent phenomena such as (ultrafast) charge transfer~\cite{Hong2014_Nature_Nanotech} and long-lived interlayer excitons~\cite{Fang2014, Palummo2015_NanoLett,Gao2017Dec,Gillen2018Apr,Torun2018Jun} have been observed, and using twist angle control, their emission quantum yield~\cite{Nayak2017_ACS_Nano} and energy~\cite{Kunstmann2018_NaturePhysics} can be tuned.
MoS$_2$/WS$_2$ heterostructures exhibit a type II band alignment~\cite{Kosmider2013_PhysRevB}. As for other material combinations, charge transfer in MoS$_2$/WS$_2$ heterostructures requires good interlayer coupling between the constituent layers, which can be achieved by annealing of the sample~\cite{Tongay2014_NanoLett}. Thereby, induced photoluminescence (PL) quenching of the monolayer emission in the heterostructure serves as an indicator for efficient charge transfer~\cite{Hong2014_Nature_Nanotech}.
Furthermore, Raman spectroscopy represents a non-invasive tool that has been widely used to characterize the quality of interfacial contact. In the low frequency regime, interlayer shear modes have been observed for MoS$_2$/WS$_2$ heterostructures with high-quality interfaces~\cite{Zhang2015_AdvMater, Okada2018_ACS_Nano, Saito2020_JapJournal, Shin2022_Nanomaterials}. However, due to the enlarged lattice mismatch for larger twist angles, lack of well-defined interlayer atomic registry and hence the missing of a restoring force, the interlayer shear mode is only expected for stacking angles of  0$^\circ°$ or 60$^\circ°$~\cite{Holler2020_ApplPhysLett, Lui2015_PhysRevB}. A more comprehensive picture, that is also applicable to twisted heterostructures, is obtained by evaluating the behavior of the in- (E$^1_{2g}$) and out-of-plane (A$_{1g}$) high-frequency Raman modes. Indicating good interlayer contact, the stiffening of the A$_{1g}$ Raman modes~\cite{Zhou2014_ACS_Nano, Zhang2015_AdvMater}, the frequency difference A$_{1g}$-E$^1_{2g}$~\cite{Liang2014_Nanoscale, Saito2020_JapJournal, Wu2021_Nano_Research}, the A$_{1g}$ linewidth~\cite{Wu2021_Nano_Research} as well as the intensity ratio A$_{1g}$/E$^1_{2g}$~\cite{Saito2020_JapJournal} have been studied extensively for MoS$_2$/WS$_2$ heterostructures. While an enhancement of the WS$_2$ A$_{1g}$ mode in the heterostructure compared to the isolated monolayer has been observed and attributed to a strong interlayer coupling~\cite{Wu2021_Nano_Research, Saito2020_JapJournal},  a clear understanding of the microscopic origin of this enhancement is lacking, so far.  

Here we present a detailed analysis of the effect and uncover its origin using optical spectroscopy on MoS$_2$/WS$_2$ heterostructures with different interlayer twist angles.
In general, systematic studies of indicators for high-quality interfacial contact, as well as twist-angle-dependent optical signatures, are hampered by the fact that challenges of controlling interlayer twist angle and coupling during sample fabrication typically lead to large variance from sample to sample.  Using a hybrid fabrication approach combining exfoliated and high-quality chemical-vapor-deposition-grown (CVD) TMDC monolayers~\cite{Shree2019_2D_Mater, George2019_JPhysMater}, we are able to produce a large set of different twist angles in a single preparation process with well-defined parameters. We find a pronounced, selective enhancement of the WS$_2$ A$_{1g}$ mode in heterostructures compared to isolated monolayers. This enhancement strongly depends on the interlayer twist. Remarkably, it is observed for both Stokes and anti-Stokes processes, indicating its nonresonant origin. The latter is further examined by contrasting the nonresonant measurements with wavelength- and temperature-dependent Raman experiments, that reveal a typical resonance behavior enhancing all WS$_2$ modes. The nonresonant, mode-specific Raman enhancement is confirmed using ab-initio calculations of the nonresonant Raman intensities based on density functional theory (DFT).

\section{Results and discussion}

\begin{figure} [h]
    \centering
    \includegraphics[width=1.0\linewidth]{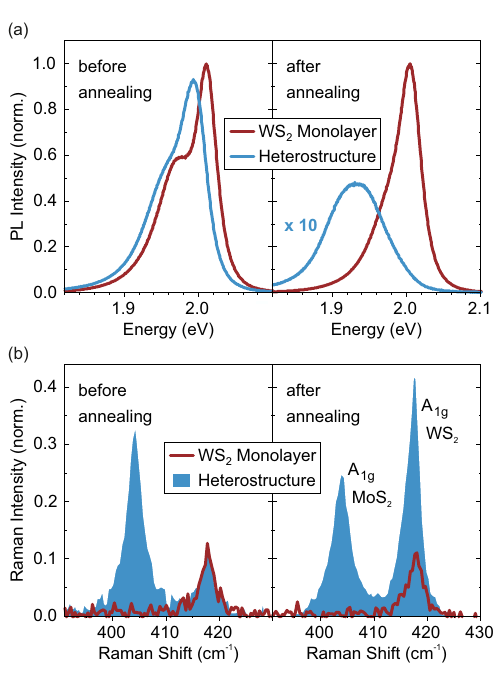}
    \caption{(a) Room temperature PL spectra of a MoS$_2$/WS$_2$ heterostructure. After annealing, PL quenching of the WS$_2$ A exciton emission occurs in the heterostructure region, indicating good contact between the constituent layers. Corresponding PL maps are shown in Supplementary Information~S1. (b) Initially, Raman spectra of the same spots do not show intensity differences for the WS$_2$ A$_{1g}$ mode in the monolayer (red line) and heterostructure region (blue area). However, the WS$_2$ A$_{1g}$ mode is enhanced after annealing. Raman spectra are normalized to the intensity of the Si phonon on the bare Si/SiO$_2$ substrate. Both Raman and PL spectra were acquired with an excitation wavelength of 532\,nm.}
    \label{Panel1_Raman_Enhancement}
\end{figure}

The impact of the interfacial quality in MoS$_2$/WS$_2$ heterostructures on PL and Raman features is illustrated in Figure~\ref{Panel1_Raman_Enhancement}. Before the sample was annealed, almost equal PL intensities of the WS$_2$ A exciton emission in the heterostructure and isolated WS$_2$ monolayer regions indicate only weak interaction between the heterostructure's individual layers (Fig.~\ref{Panel1_Raman_Enhancement}a). The small spectral shift observed for the A exciton emission can be attributed to the different dielectric environments provided by the SiO$_2$ and MoS$_2$, respectively. In contrast, a pronounced quenching of the WS$_2$ emission is observed in the heterostructure region after annealing. Interestingly, Raman spectra measured at the same sample position also show  significant differences (Fig.~\ref{Panel1_Raman_Enhancement}b). Whereas initially the WS$_2$ out-of-plane A$_{1g}$ Raman mode intensity does not differ between the WS$_2$ monolayer and the heterostructure region, a strong enhancement of this mode occurs after annealing. This suggests that the enhancement of the WS$_2$ A$_{1g}$ mode serves as an additional indicator for high-quality MoS$_2$/WS$_2$ heterostructure interfaces. In the following, we further characterize the effect and reveal its microscopic origin.

\begin{figure*} [t]
    \centering
    \includegraphics[width=1.0\linewidth]{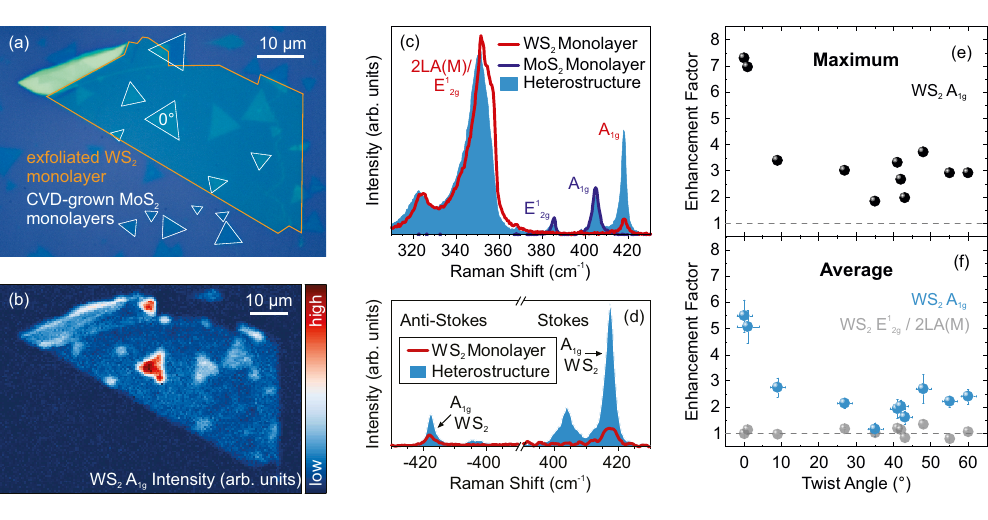}
    \caption{(a) Optical microscope image of a sample consisting of CVD-grown MoS$_2$ monolayers which were covered with a large exfoliated WS$_2$ flake. This yields different heterostructures with various twist angles in the same sample. Some CVD-grown monolayers were marked for illustration purposes. (b) Raman scan (532\,nm excitation) of the same sample showing the intensity of the WS$_2$ A$_{1g}$ Raman mode. (c) Averaged Raman spectra of the marked heterostructure with 0$^\circ°$ twist angle. Compared to other Raman modes, the WS$_2$ A$_{1g}$ is clearly enhanced. Averaged spectra for all heterostructure regions are shown in Supplementary Information~S3. (d) The enhancement appears on the Stokes and Anti-Stokes side of the Raman spectrum. The spectrum was measured in the center of the 0$^\circ°$ twist angle heterostructure. (e,f) Maximum and average enhancement factors of the WS$_2$ A$_{1g}$ mode obtained for heterostructures with different twist angles. For comparison, 2LA(M)/E$^1_{2g}$ average enhancement factors are included, underlining the mode selectivity of the enhancement process.  Error bars represent statistical uncertainties determined from all selected individual WS$_2$ spectra.}
    \label{Panel_2_Raman_Enhancement}
\end{figure*}

Figure~\ref{Panel_2_Raman_Enhancement}a shows MoS$_2$/WS$_2$ heterostructures that were fabricated by placing a large exfoliated WS$_2$ monolayer on top of CVD-grown MoS$_2$ monolayers (see methods). This yields the advantage of creating several heterostructures with various twist angles within one preparation step. Here, determination of the interlayer twist angle is enabled by optical microscopy. We analyzed the angle between a CVD triangle edge and the long straight WS$_2$ monolayer edge, both assumed to follow high-symmetry crystal directions. Given the fact that slight variations of preparation parameters can alter a heterostructure's optoelectronic properties, this method also facilitates a direct comparison between the resulting heterostructures since all structures were produced under the same lab conditions and faced the same annealing conditions. Improvement of the interfacial quality caused by annealing was verified by observation of PL quenching (Supplementary Information~S2). 

The sample was mapped using nonresonant Raman spectroscopy at 532\,nm excitation. The corresponding Raman intensity map of the WS$_2$ A$_{1g}$ mode reveals a higher intensity in almost all heterostructure regions compared to the WS$_2$ monolayer, which proves the robustness of the effect (Fig.~\ref{Panel_2_Raman_Enhancement}b). Interestingly, a significantly stronger increase is found for stacking angles close to 0$^\circ°$. Raman spectra belonging to the 0$^\circ°$ heterostructure marked in (a) were extracted from the scan data, and then summed and divided by the number of selected spectra. The resulting averaged spectrum is depicted in Figure~\ref{Panel_2_Raman_Enhancement}c and compared to WS$_2$ and MoS$_2$ monolayer spectra that were obtained in the same way. Its characteristic features are the out-of-plane A$_{1g}$ and in-plane E$^1_{2g}$ optical modes of both WS$_2$ and MoS$_2$ as well as the longitudinal acoustic 2LA(M) mode of WS$_2$ \cite{Berkdemir2013_Scientific_Reports}. For simplicity, we consider the WS$_2$ 2LA(M) (352\,cm$^{-1}$) and  E$^1_{2g}$ (356\,cm$^{-1}$) mode as one combined peak.

In contrast to the WS$_2$ A$_{1g}$ mode, the MoS$_2$ mode intensities as well as the combined WS$_2$ 2LA(M)/E$^1_{2g}$  peak do not differ from their monolayers' counterparts. In the following, we thus primarily focus on the WS$_2$ A$_{1g}$ mode behavior. Figure \ref{Panel_2_Raman_Enhancement}d shows that the effect is also apparent on the anti-Stokes side of the Raman spectrum. Together with the absence of enhancement of the WS$_2$ combined 2LA(M)/E$^1_{2g}$ peak, this implies that the detected enhancement does not result from a resonant Raman process, as a laser excitation energy matching a WS$_2$ monolayer optical transition (incoming resonance) would affect all WS$_2$ modes marked in the spectrum \cite{Corro2016_Nano_Lett}.    

\begin{figure*} [t]
    \centering
    \includegraphics[width=1.0\linewidth]{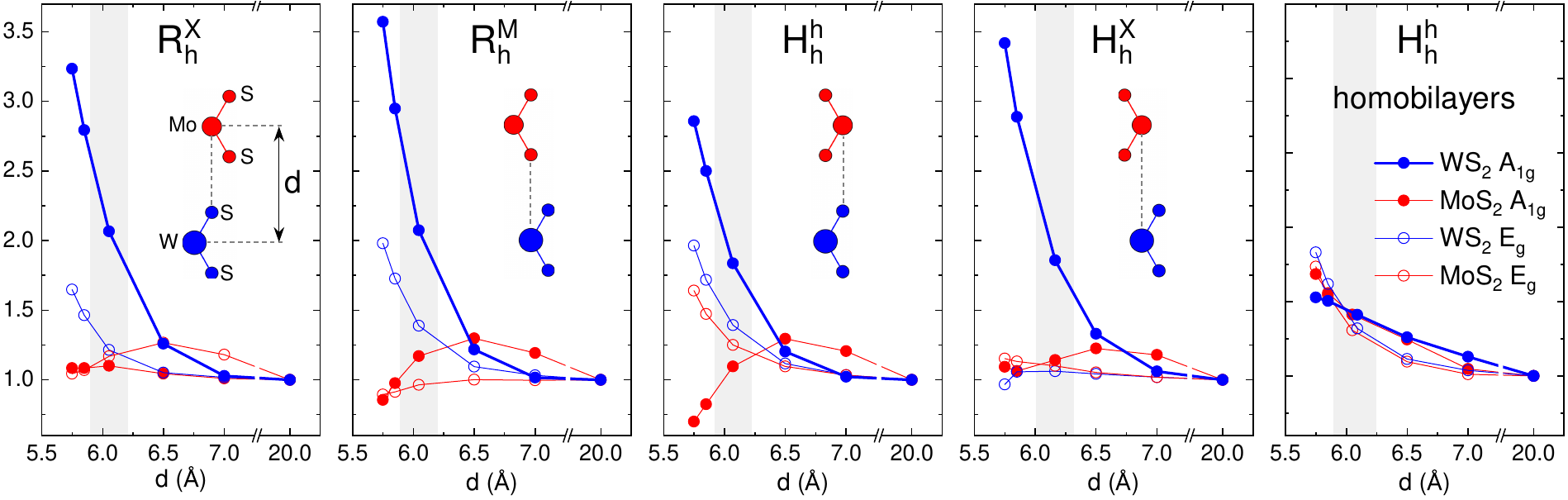}
    \caption{Calculated Enhancement factors of WS$_2$ and MoS$_2$ (blue and red) A$_{1g}$ and E$_g$ (solid and hollow points) Raman modes in WS$_2$/MoS$_2$ heterostructures and $H_h^h$ homobilayers as a function of the interlayer distance. The range of equilibrium interlayer distances for each heterostructure is marked by a gray vertical bar. The insets depict the side views of atomic structures of the heterobilayers, where larger (smaller) blue (red) circles represent metal (sulfur) atoms in WS$_2$ (MoS$_2$) layer, and the gray dashed lines connect atoms that are vertically aligned in the given stackings.}
    \label{theory}
\end{figure*}

For further twist angle-dependent analysis, Raman spectra of the different heterostructures and the surrounding WS$_2$ monolayer were selected from the Raman scan shown in Figure~\ref{Panel_2_Raman_Enhancement}b. The WS$_2$ A$_{1g}$ mode intensity of every spectrum was then obtained by numerical integration over the energetic region of interest. The maximum enhancement factor (Fig.~\ref{Panel_2_Raman_Enhancement}e) is defined as $I_{HS}/I_{\overline{ML}}$, where $I_{HS}$ is the highest WS$_2$ A$_{1g}$ mode intensity value determined in the respective heterostructure and $I_{\overline{ML}}$ the average monolayer WS$_2$ A$_{1g}$ mode intensity. In Figure~\ref{Panel_2_Raman_Enhancement}f, the average enhancement factor $I_{\overline{HS}}/I_{\overline{ML}}$ for various heterostructures is depicted in dependence of the twist angle. We observe a reduced enhancement for twist angles deviating from 0$^\circ°$, reaching no observable enhancement close to 30$^\circ°$, followed by a slight increase up to 60$^\circ°$. This trend resembles previously reported results~\cite{Wu2021_Nano_Research} for twisted WS$_2$/MoS$_2$ heterostructures where both layers were obtained from CVD growth. Given that the interlayer distance in twisted WS$_2$/MoS$_2$ depends on the twist angle with a smallest layer separation at 0$^\circ°$ and 60$^\circ°$~\cite{Wu2021_Nano_Research, Yang2018_Nanoscale, Tebyetekerwa2021_Cell_Reports}, these results suggest that the enhancement is sensitive to the interlayer distance. Moreover, we assume that the different stacking orders of the involved atoms, namely R-type (0$^\circ°$) and H-type (60$^\circ°$) stacking \cite{Yang2018_Nanoscale, Rosenberger2020_ACS_Nano}, explain the different enhancement factors at comparable interlayer distance. Average enhancement factors for the WS$_2$ 2LA(M)/ E$^1_{2g}$ peaks occurring in the same heterostructures were calculated accordingly. To this end, we integrated over the combined WS$_2$ 2LA(M)/ E$^1_{2g}$ peak and the shoulder at 325\,cm$^{-1}$, since the phonon modes overlap in this region. Contrary to the WS$_2$ A$_{1g}$ mode, the WS$_2$ 2LA(M)/E$^1_{2g}$ peaks do not show any pronounced intensity increase which underlines the mode selectivity of the enhancement process. 

We note that the enhancement is not limited to MoS$_2$/WS$_2$ heterostructures consisting of both exfoliated and CVD-grown monolayers, but also occurs in purely exfoliated samples (Supplementary Information S4). Interestingly, we also observe an enhancement of the WS$_2$ A$_{1g}$ mode in MoSe$_2$/WS$_2$ heterostructures (Supplementary Information S5).

We complement the nonresonant Raman measurements in MoS$_2$/WS$_2$ heterostructures with DFT calculations. As MoS$_2$ and WS$_2$ have nearly identical lattice constants, their heterostructures exhibit significant atomic reconstructions. Instead of moiré patterns, large domains with high-symmetry stackings emerge, as has been observed in MoS$_2$/WS$_2$ and MoSe$_2$/WSe$_2$ heterostructures \cite{Rosenberger2020_ACS_Nano,Holler2020_ApplPhysLett,Wang2022}. In the following, the stacking configurations are labeled according to Yu \textit{et al}.~\cite{Yu2017_ScienceAdv, Yu2018_2DMater}. Samples with twist angles near 0$^\circ°$ (60$^\circ°$) consist mostly of $R^X_h$ and $R^M_h$ ($H^h_h$ and $H^X_h$) domains, with characteristic vertical alignment of atoms as depicted by insets in Figure~\ref{theory}. We neglect the $R^h_h$, $H^M_h$ and intermediate stacking registries, as they cover a minor part of the sample and their contribution to the Raman signal is negligible. The stackings are modeled within primitive cells and start from finding their equilibrium interlayer distances $d_{eq}$, which is defined as the vertical distance between Mo and W atomic planes. Then we calculate the Raman intensities $I=I_{xx}+I_{yy}$ of the MoS$_2$ and WS$_2$ A$_{1g}$ and E$_g$ modes for a series of interlayer distances $d$. Other components of the Raman tensor are omitted, since they are not probed in the experimental back-scattering measurements. Figure~\ref{theory} presents the enhancement factors (EFs) of these modes in function of $d$, defined as:
\begin{equation}
    EF=\frac{I(d)}{I(d_{1L})}.
    \label{EF}
\end{equation}
Here, $I(d_{1L})$ is the Raman intensity of the respective mode in the monolayer.
Distinctly, in all the considered stackings, the enhancement factor of the WS$_2$ A$_{1g}$ mode increases when the layers get closer to each other, and reaches $\approx 2$ at $d_{eq}$. When $d$ is further decreased by $\approx 10\%$ with respect to $d_{eq}$, the enhancement factor of the WS$_2$ A$_{1g}$ mode rapidly grows, exceeding $\approx 3.5$ in $R^M_h$. It qualitatively agrees with experimental findings, and supports the hypothesis that annealing leads to a reduction of interlayer distance in MoS$_2$/WS$_2$ samples, which is revealed by the enhancement of the WS$_2$ A$_{1g}$ Raman mode intensity.
The enhancement factors of other modes exhibit various dependence on $d$, depending on the stacking. However, their values at $d_{eq}$ and lower do not exceed 2, which is in line with experimental observations. For comparison, we calculated and plotted the enhancement factors of the discussed modes for MoS$_2$ and WS$_2$ homobilayers in $H^h_h$ stacking, which corresponds to their bulk crystals. Interestingly, intensities of all the modes are enhanced with similar trends, but to a lower extent than in MoS$_2$/WS$_2$ heterostructures. 

In order to understand the selective enhancement of the aforementioned Raman modes, let us look at the quantities that contribute to the off-resonant Raman intensity calculated within Placzek's approximation~\cite{placzek1934rayleigh}. For the back-scattering measurement geometry, the Raman intensity (for light scattering between Cartesian directions $i$ and $j$) of the phonon mode $\nu$ is
\begin{equation}
    I_{ij}^\nu \sim \left| \sum\limits_{k,n}  \alpha'_{ijkn} u_{kn}^\nu \right|^2
    \label{raman_formula}
\end{equation}
with $\alpha'_{ijkn}= \frac{\partial \chi_{ij}}{\partial \tau_{kn}}$ the atom-resolved Raman polarizability defined as derivative of $\chi_{ij}$, the macroscopic dielectric susceptibility, with respect to $\tau_{kn}$, the displacement of the $n$th atom in the Cartesian direction $k$. The eigen-displacement of phonon mode $\nu$ of atom $n$ in direction $k$ is denoted by $u_{kn}$. The prefactor cancels out in the calculation of EFs, and thus is omitted in Equation~\ref{raman_formula} for brevity.

Let us discuss the $R^X_h$ heterostructure as a representative case, and WS$_2$ and MoS$_2$ $H^h_h$ homobilayers as reference. The atom-resolved Raman polarizabilities and eigen-displacements of the WS$_2$ and MoS$_2$ A$_{1g}$ and E$_g$ phonon modes are presented in Figure~S6 as a function of interlayer distance. We focus on the non-zero components that contribute to $I_{xx}$: From symmetry it follows that $I_{yy}=I_{xx}$ and $I_{xy}=0$ for the A$_{1g}$ out-of-plane modes where $u_{xn}=u_{yn}=0$. For the E$_{g}$ in-plane modes, $u_{zn}=0$. As discernible in Figure~S6, the eigen-displacements and Raman polarizabilites are more sensitive to the interlayer distance for A$_{1g}$ modes than for E$_g$ modes. Furthermore, for the A$_{1g}$ modes, the interlayer dependence of the $u_z$ vibration amplitude is more pronounced for the sulfur atoms of the WS$_2$ layer than for those of the MoS$_2$ layer. 

But the decisive point for the selective enhancement of the WS$_2$ A$_{1g}$ mode is the relative phase of the sulfur atom vibrations on both layers. At large interlayer distance, the vibrations are exclusively localized on the WS$_2$ or MoS$_2$ layer, respectively. However, with decreasing interlayer distance, the interlayer coupling leads to a ``minority vibration'' of the sulfur atoms on the adjacent layer. At equilibrium, the amplitude ratio is $1:7$ and rapidly 
increases upon further squeezing of the layers. As can be seen from the orientation of the triangles in Figure S6, for the WS$_2$ A$_{1g}$ mode, the vibrations of sulfur atoms (albeit with very different amplitudes) are in phase, while they are out of phase for the MoS$_2$ A$_{1g}$ mode. In phase vibration leads to two contributions with the same sign in the summation of Equation~\ref{raman_formula} and hence to a visible enhancement of the WS$_2$ mode while out of phase vibration leads to a partial cancellation in the summation and a reduction of the MoS$_2$ mode intensity.
We verified this effect by finite-displacement calculations of the Raman tensor: if only the sulfur atoms on the WS$_2$ (MoS$_2$) layer are displaced, the intensities of both A$_{1g}$ modes are slightly enhanced (as compared to the isolated single layer). Including the constructive (destructive) interference from the ``minority vibration'', one reproduces the experimentally observed and calculated selective mode enhancement of the WS$_2$ mode. The strong distance dependence of the ``minority'' amplitude explains then the observed strong distance dependence of the enhancement factor for the WS$_2$ A$_{1g}$ mode.

It should be noted that our theoretical reasoning is at the off-resonance limit and neglects resonant effects, which can be significant in layered structures, as shown in \cite{Garrity2024,Reichardt2020,Nalabothula2025}. However, resonant Raman calculations from first principles are beyond the scope of this study.

In order to gain further experimental insight into the WS$_2$ A$_{1g}$ mode’s behavior and to separate the previously described eigenmode effect from resonant effects, we also performed near-resonant and resonant Raman measurements. 
Resonant conditions were achieved by (i) sweeping the laser excitation energy and (ii) tuning the WS$_2$ and MoS$_2$ bandgaps via temperature variation. 

\begin{figure} [b]
    \centering
    \includegraphics[width=1.0\linewidth]{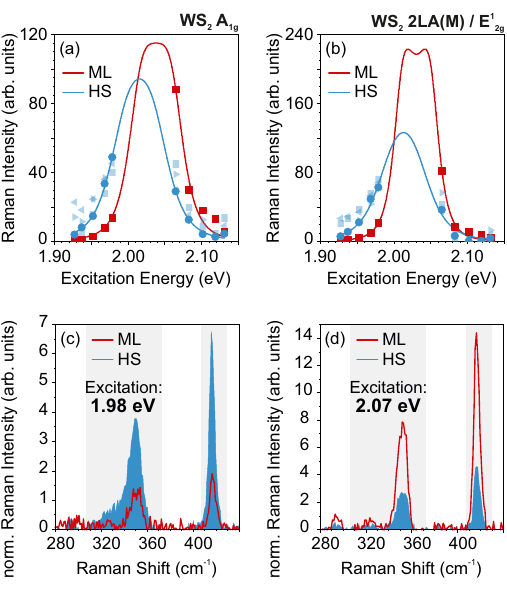}
    \caption{(a,b) Integrated Raman intensities for the WS$_2$ out-of-plane and in-plane Raman modes in a WS$_2$ monolayer (red) and several heterostructures (blue). Raman spectra were obtained from various spots on different heterostructures (symbols) shown in Figure~\ref{Panel_2_Raman_Enhancement}a, triangles indicate data originating from the 0$^\circ°$ heterostructure. Data points marked with square and circle represent measurements from the 9$^\circ°$ and 55$^\circ°$ heterostructure, respectively. A laser of tunable wavelength was used for excitation. The lines are fitted resonance curves for the isolated monolayer and the 55$^\circ°$ heterostructure. (c,d) Exemplary Raman spectra excited below (1.98\,eV) and above (2.07\,eV) the WS$_2$ A exciton energy. The energy range used for intensity determination via numerical integration is highlighted in gray.}
    \label{Panel4_Raman_Enhancement}
\end{figure}

For the former, the laser excitation wavelength was tuned below and above the WS$_2$ A exciton resonance energy, using a Radiant Dye laser (see methods). Spectra at various excitation energies were taken on the WS$_2$ monolayer and on different heterostructures of the sample shown in Figure~\ref{Panel_2_Raman_Enhancement}a. After PL background subtraction, all spectra were normalized to the Si phonon intensity. Intensities of the Raman modes of interest were obtained by numerical integration. The resulting phonon intensities are depicted in Figures~\ref{Panel4_Raman_Enhancement}a and~\ref{Panel4_Raman_Enhancement}b, exemplary Raman spectra above and below the WS$_2$ A exciton resonance are shown in Figures~\ref{Panel4_Raman_Enhancement}c and~\ref{Panel4_Raman_Enhancement}d.

The data points (symbols) constitute the tails of typical Raman resonance curves~\cite{Corro2016_Nano_Lett}. Below the A exciton resonance (i.e., at lower excitation energies), larger intensities are detected for both the out-of-plane $A_{1g}$ Raman mode and the combined 2LA(M)/ E$^1_{2g}$ peak at all heterostructure spots compared to the isolated monolayer. In contrast, higher intensities occur for the monolayer's modes above the WS$_2$ A exciton energy. With increasing deviation from the WS$_2$ A exciton energy, the intensities of monolayer and heterostructure converge. At the resonance, no data points could be recorded since the available dyes did not allow stable laser emission in the relevant energy range. Resonance profiles (lines) were subsequently generated by fitting the Raman intensities using third-order perturbation theory~\cite{Jorio2011, Tan2019_Springer, Garrity2024}:

\begin{equation}
    I_R(\omega_{ph},E_l) \propto  \left| \frac{M_{op}^2 \cdot M_{ep}}{(E_l - E_a +i\gamma_a)(E_l-\hbar\omega_{ph}-E_a+i\gamma_a)} \right|^2 
\end{equation}

We combined the matrix elements $M_{op}$ (electric dipole transition) and $M_{ep}$ (electron-phonon interaction) into one fitting parameter. Additional parameters include the energy $E_a$ of the intermediate resonant state, which is the WS$_2$ A exciton, and the decay rates $\gamma_a$ of the incoming and outgoing resonance. As common, the same value $\gamma_a$ was assumed for both resonance events~\cite{Jorio2011}. $E_l$ is the laser excitation energy and $\hbar\omega_{ph}$ the respective phonon energy. In general, the energy of the A exciton in the heterostructure regions is redshifted relative to that in the monolayer~\cite{Rigosi2015_NanoLett}. The exemplary resonance profiles (lines) for one heterostructure spot and the isolated WS$_2$ monolayer clearly reflect this trend. Compared to the monolayer, the resonant condition in the heterostructure regions already sets in at lower excitation energies. The intensities of both out-of-plane and in-plane modes follow the resonance curve for all heterostructure positions. We attribute the rather broad peak for the A$_{1g}$ intensity profile in the monolayer region to an overlap of incoming and outgoing resonance, which is more distinctly resolved for the in-plane modes. The data were obtained for heterostructures with 0$^\circ°$, 9$^\circ°$ and 55$^\circ°$ twist angle. Interestingly, in none of these heterostructures does the 2LA(M)/ E$^1_{2g}$ peak show any distinct enhancement at 532\,nm excitation (Supplementary Information S3). Therefore, the observed increase of both out-of-plane and in-plane Raman modes - which is typical resonant behavior as previously observed in WS$_2$ monolayers~\cite{Corro2016_Nano_Lett} - provides evidence that different enhancement mechanisms dominate under near-resonant and nonresonant excitation conditions.
 
\begin{figure} [h]
    \centering
    \includegraphics[width=1.0\linewidth]{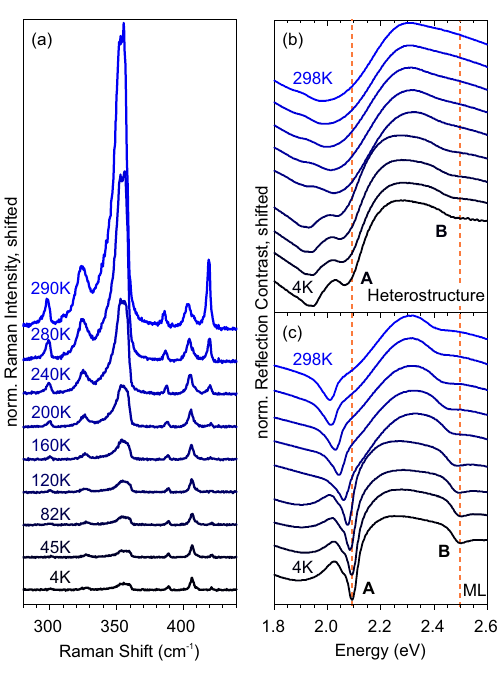}
    \caption{(a) Temperature-dependent Raman spectra of a  MoS$_2$/WS$_2$ heterostructure at 532~nm excitation. All spectra are normalized to the Si phonon.  Raman modes of interest are the WS$_2$ A$_{1g}$ mode at 420\,cm$^{-1}$ and the combined 2LA(M)/ E$^1_{2g}$ peak at 355\,cm$^{-1}$.  (b,c) Low-temperature white-light reflectance contrast (RC) measurements of the same sample. $RC=(R_{Sample}-R_{Ref})/R_{Ref}$ was used for normalization, with $R_{Ref}$ being the reflectance of the bare Si/SiO$_2$ substrate. Furthermore, all spectra were normalized to the minimum of the WS${_2}$ A exciton for better illustration of energetic shifts. The dotted line indicates the WS$_2$ monolayer's A and B exciton energies at 4\,K.}
    \label{Panel3_Raman_Enhancement}
\end{figure}

Further temperature-dependent measurements were performed on the sample introduced in Figure~S4. Figure \ref{Panel3_Raman_Enhancement}a shows temperature-dependent Raman spectra at 532\,nm laser excitation. All spectra were normalized to the Si phonon intensity of each individual spectrum. As the temperature rises, both the WS$_2$ A$_{1g}$ and the combined WS$_2$ 2LA(M)/ E$^1_{2g}$ mode intensity increase, with the effect starting to become pronounced at about 160\,K. Similar temperature-dependent behavior was reported for WS$_2$ monolayers at 514.5\,nm (2.41\,eV) excitation and explained by resonance with the WS$_2$ B exciton \cite{Huang2016_Scientific_Reports}. 

Despite the lower excitation energy in our experiment (2.33\,eV), we still observe the onset of a resonant process due to temperature-dependent modifications in the WS$_2$ bandstructure in both monolayer and heterostructure. To illustrate this effect, we performed white-light reflectance measurements on the MoS$_2$/WS$_2$ heterostructure and the isolated WS$_2$ monolayer (Fig.~\ref{Panel3_Raman_Enhancement}b,c). Here, several relevant aspects are apparent: First, the characteristic absorption features of WS$_2$ A and B exciton are broader and redshifted in the heterostructure compared to the monolayer; and second, in both monolayer and heterostructure a broadening and redshift of the absorption dips occur with increasing temperature. Thus, resonance with the WS$_2$ B exciton is established in the heterostructure region close to room temperature. Consequently, we observe a pronounced enhancement of both in- and out-of-plane Raman modes compared to the isolated monolayer (Supplementary Information~S4). Remarkably, the WS$_2$ A$_{1g}$ mode intensity is more affected than the combined 2LA(M)/~E$^1_{2g}$ mode intensity. Analogous to the the calculations for the heterostructures presented in Figure~\ref{Panel_2_Raman_Enhancement}, we extract from the room temperature Raman scan (Fig.~S4) an average enhancement factor of 4.55~$\pm$~0.46 for the WS$_2$ A$_{1g}$ mode and an average enhancement factor of 1.487~$\pm$~0.045 for the combined WS$_2$ 2LA(M)/ E$^1_{2g}$ peak. This is in contrast to previously reported resonant Raman profiles for WS$_2$ monolayers, where the enhancement at the WS$_2$ B exciton energy is slightly smaller for the WS$_2$ A$_{1g}$ than for the WS$_2$ 2LA(M) and E$^1_{2g}$ modes~\cite{Corro2016_Nano_Lett}, which indicates that we observe a superposition between resonant enhancement and the mode-selective effect occurring in MoS$_2$/WS$_2$ heterostructures. We note that, with both WS$_2$ peaks enhanced at room temperature compared to the isolated monolayer, the sample's spectral behavior differs from that of the zero-degree twist angle heterostructure spectrum shown in Figure~\ref{Panel_2_Raman_Enhancement}c. However, the comparability of various samples with regard to their resonance energy is complex due to multiple factors. The exciton binding energy - and hence the A exciton resonance - is sensitive to dielectric screening which is governed by the interlayer distance~\cite{Florian2018_NanoLett}. The latter in turn is highly affected by contaminants introduced during sample preparation such as hydrocarbons~\cite{Haigh2012_NatMater} and PDMS residues~\cite{Jain2018_Nanotechnology}. Furthermore, band gaps can be modified by strain occurring in the heterostructure region~\cite{He2013_NanoLett, Conley2013_NanoLett, Plechinger2015_2D_Mater, He2016_ApplPhysLett}. Since the energetic difference between A and B exciton is expected to remain almost constant for distinct samples~\cite{Zhu2015_ScientificRep}, these factors have a direct impact on the resonance condition discussed above. 
The WS$_2$ A$_{1g}$ mode enhancement in MoS$_2$/WS$_2$ heterostructures is thus driven by a complex interplay of interlayer distance between the constituent layers and resonant conditions that depend on the individual heterostructure's properties.

In summary, we have fabricated vdW heterostructures by combining CVD grown MoS$_2$ monolayers with large exfoliated WS$_2$ monolayers. This approach allows us to achieve comparable preparation conditions and facilitates comparison between various samples. Under nonresonant conditions, we observe a twist-angle-dependent mode-selective enhancement of the WS$_2$ A$_{1g}$ mode which serves as an easily accessible indicator for high-quality interfacial contact. DFT calculations reveal that it originates from an in-phase oscillation of the MoS$_2$ sulfur atoms with those in WS$_2$ for the A$_{1g}$ mode displacement. Near-resonant Raman measurements, realized by varying the excitation wavelength or the temperature, demonstrate that the mode-selective enhancement is independent of resonant effects. Our results highlight the complex interplay of phonon modes across the van der Waals gap.

\section{Methods}
\subsection{Sample fabrication}

CVD-grown~\cite{George2019_JPhysMater} MoS$_2$ monolayers were picked up from the growth substrate by capillary-force assisted transfer \cite{Ma2017_NanoLett}. For this, we either used deionized water vapor to wet a PDMS stamp or a deionized water droplet that was carefully placed on the growth substrate prior to pick-up. MoS$_2$ monolayers were then transferred to a Si/SiO$_2$ substrate by viscoelastic stamping \cite{CastellanosGomez2014_2D_Mater}. Several MoS$_2$ monolayer triangles were covered with a large WS$_2$ monolayer, that was previously mechanically exfoliated onto PDMS, in a second transfer step. All exfoliated flakes were obtained from bulk crystals by HQ graphene. To increase interlayer contact, samples were annealed in high vacuum at 420\,K for 3\,hours.

\subsection{Optical measurements}
\subsubsection{Raman and Photoluminescence Measurements}

For PL and nonresonant Raman measurements, a continuous-wave 532~nm diode-pumped solid state laser was used for excitation. The laser beam was focused on the sample through a 100x (room temperature) or a 80x (low temperature) microscope objective. The scattered/emitted light was collected through the same objective. Mounted on a motorized xy-stage, the sample was moved with \textit{$\mu$}m precision under the fixed laser spot position. A set of three successive Bragg filters suppressed Rayleigh-scattered light in front of a spectrometer equipped with a CCD sensor operated at -70$^\circ°$C. For low-temperature Raman measurements, the sample was placed in a flow cryostat cooled with liquid helium.

For (near-)resonant Raman measurements the sample was excited with a wavelength-tunable Radiant Dye Laser, using two dyes (DCM and  R6G). With an excitation power of 1.1\,mW the laser beam was focused through an 100x microscope objective under which the sample was placed. The scattered light was collected using the same objective and guided into a Horiba T64000 micro-spectrometer with triple-grating configuration. The Raman signal was dispersed by a grating with 900 grooves per mm, and a CCD sensor was used for detection. All (near-)resonant Raman measurements were performed at room temperature.
\newline

\subsubsection{White Light Reflectance Measurements}

White-light reflectance contrast measurements were performed using a quartz tungsten halogen lamp. A collimated beam was focused on the sample by a 80x objective. The reflected light was collected through the same objective and was guided into a spectrometer equipped with a CCD detector. Enhanced spatial resolution was achieved by a spatial-filtering module in the detection path (see \cite{Deb2024_Nature_Commun} for details). The sample was mounted in a flow cryostat cooled with liquid helium. 

\subsection{Theoretical calculations}
Density functional theory calculations were performed with the Quantum Espresso package \cite{QE-2009,QE-2017} within local density approximation (LDA) to exchange-correlation functional and norm-conserving scalar-relativistic pseudopotentials of version 1.2 \cite{Hamann-2013}. The energy cut-off for the wave functions was 90 Ry. The Brillouin zone was sampled with a $\Gamma$-centered Monkhorst-Pack grid of 12$\times$12$\times$1 k-points. The in-plane and out-of-plane lattice constants of 3.1270\,\AA~and 40\,\AA~were used for all the structures. The former represents the average of experimental values for WS$_2$ and MoS$_2$. The latter yields an effective vacuum distance between repeated hetero-layers of 20\,\AA, which is sufficient to reproduce the phonons of isolated hetero-layers, as verified by convergence tests. The atomic positions were optimized until all the forces acting on the atoms were less than $10^{-6}$ Ry/Bohr. When varying the interlayer distance, positions of metal atoms were kept fixed during the optimization. The phonons at $\Gamma$ were calculated using density functional perturbation theory (DFPT). The nonresonant Raman intensities were evaluated with DFPT, using the implementation of Lazzeri and Mauri\cite{Lazerri-2003}. For the proper treatment of 2D boundary conditions, the Coulomb cut-off technique was used in all calculations \cite{Sohier2107}.

\section{Acknowledgements}
T.K. acknowledges financial support by the DFG \emph{via} the following grants: SFB1477 (project No. 441234705) and SPP2244 (project No. 443361515). 
L.W. acknowledges funding by the FNR Luxembourg through project C22/MS/17415967/ExcPhon. The DFT calculations were carried out using the HPC facilities of the University of Luxembourg.
PK and OG acknowledge the DFG for funding (KU4034 2-1).
Z.G., A.G. and AT acknowledge financial support of the DFG through SPP2244 "2DMP" (Projects TU149/13-1 and TU149/21-1) and CRC 1375 "NOA" (Project B2).

\maketitle
\onecolumngrid
\newpage
\renewcommand{\thefigure}{S\arabic{figure}}
\renewcommand{\thetable}{S\arabic{table}}
\setcounter{figure}{0}
\setcounter{table}{0}
\section{Supplementary experimental data}

\begin{figure*} [h]
	\centering
	\includegraphics[width=1.0\textwidth]{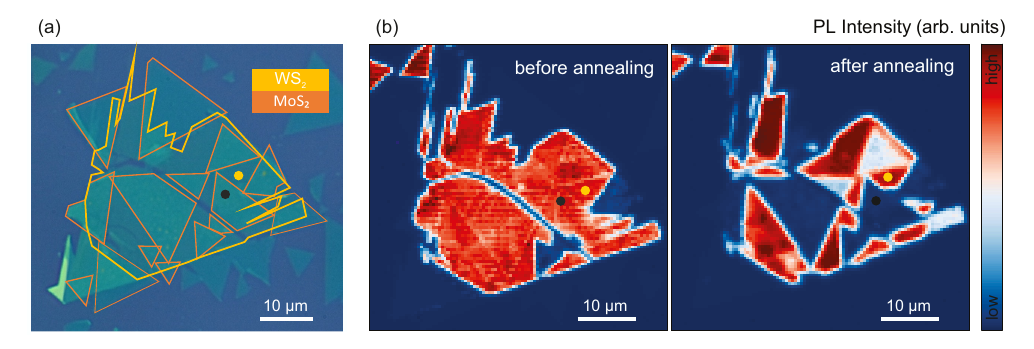}
	\caption{(a) Optical Microscope image of the MoS$_2$/WS$_2$ sample on which the spectra shown in Figure 1 (main text) were measured. CVD-grown MoS$_2$ monolayers were covered with a large exfoliated WS$_2$ monolayer, yielding several heterostructure regions. The dots indicate the measurement positions of the heterostructure (black) and WS$_2$ monolayer (yellow) spectra presented in the main text. (b) PL maps (acquired with an excitation wavelength of 532\,nm) show the spatial distribution of the WS$_2$ A exciton emission. PL quenching is observed after annealing, indicating an improvement of the heterostructure's interface.}
	\label{Supplement1}
\end{figure*}

\begin{figure*} [h]
	\centering
	\includegraphics[width=1.0\textwidth]{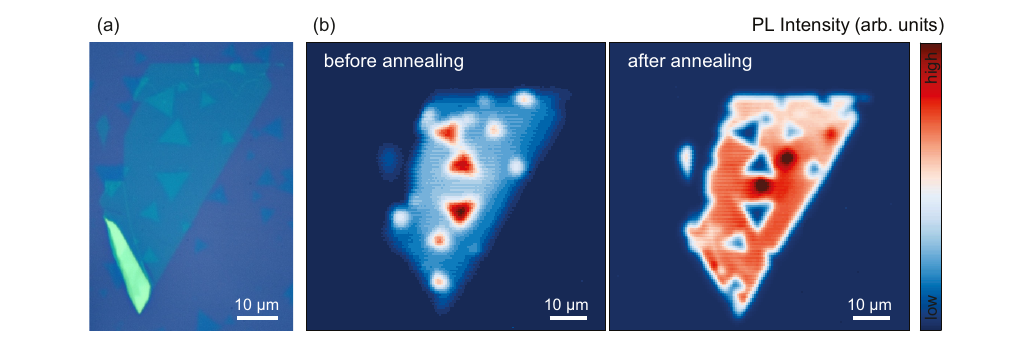}
	\caption{(a) Optical microscope image and (b) PL maps of the MoS$_2$/WS$_2$ sample introduced in Figure 2 in the main text. Good interfacial contact between the heterostructures' layers was established by annealing and verified by the observation of quenching of the WS$_2$ A-Exciton emission.  A 532\,nm laser was used for excitation.}
	\label{Supplement2}
\end{figure*}

\begin{figure*} [h]
	\centering
	\includegraphics[width=1.0\textwidth]{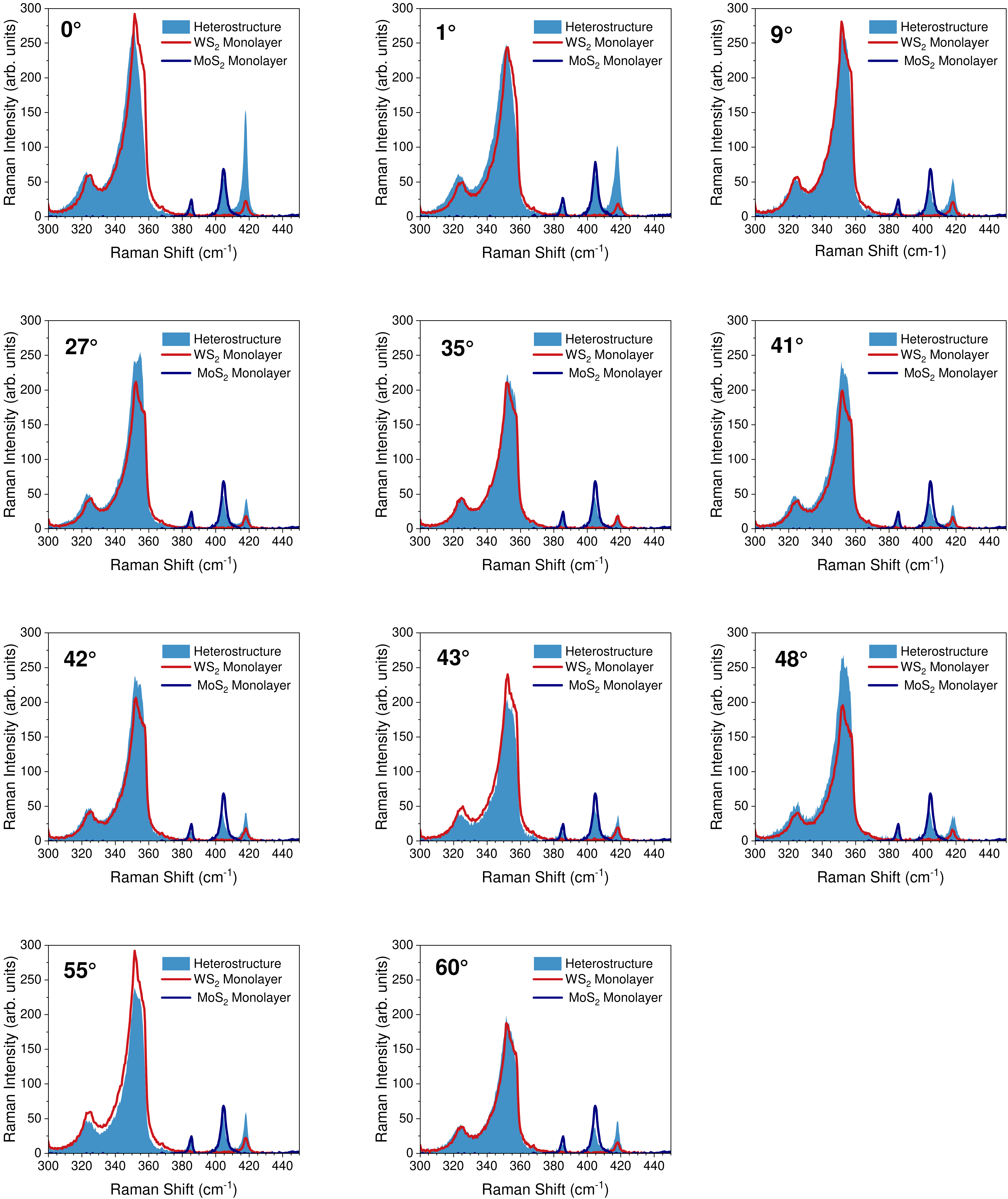}
	\caption{All averaged Raman spectra (532\,nm excitation wavelength) for the heterostructures in the sample presented in the main text (Figure~2).}
	\label{all_Raman_spectra}
\end{figure*}

\begin{figure*} [h]
	\centering
	\includegraphics[width=0.91\textwidth]{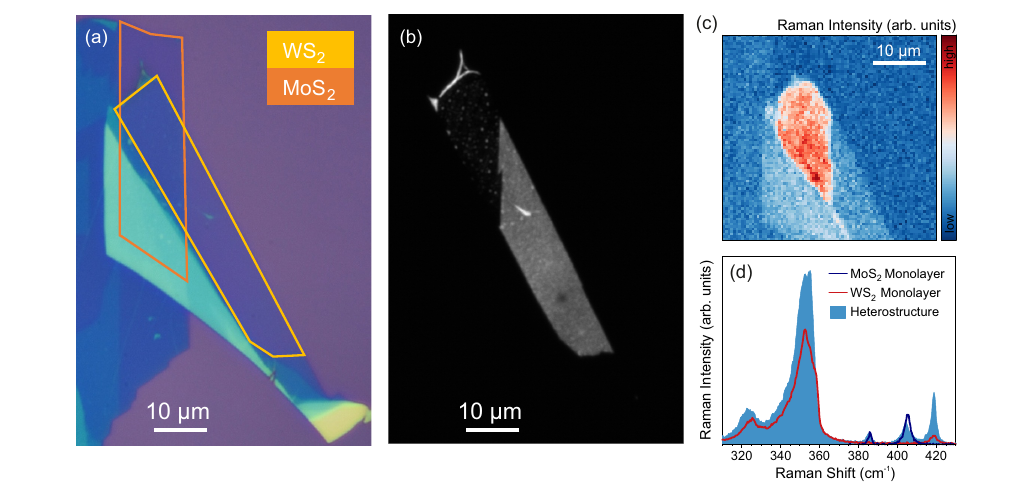}
	\caption{(a) Optical microscope image of a MoS$_2$/WS$_2$ heterostructure where both layers were obtained by exfoliation from a bulk crystal. (b) The fluorescence microscope shows quenching in the heterostructure region where an enhancement of the WS$_2$ A$_{1g}$ mode is observed in the corresponding Raman map (c). (d) Room temperature averaged Raman spectra of MoS$_2$ monolayer, WS$_2$ monolayer and heterostructure. Raman spectra were measured using an excitation wavelength of 532\,nm.}
	\label{Supplement5}
\end{figure*}

\begin{figure*} [h]
	\centering
	\includegraphics[width=0.95\textwidth]{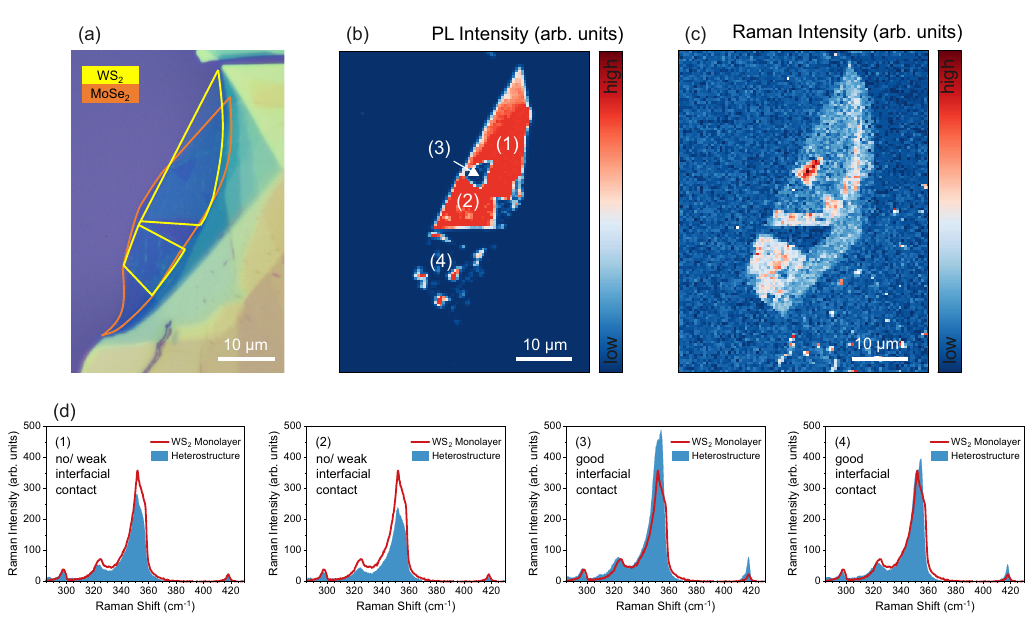}
	\caption{Characterization of a MoSe$_2$/WS$_2$ heterostructure. (a) Optical microscope image of the sample, consisting of an exfoliated MoSe$_2$ monolayer (bottom) and an exfoliated WS$_2$ monolayer (top). (b) PL map showing the WS$_2$ A exciton emission after preparation. While some heterostructure regions ((1),(2)) lack good interfacial contact, others already show PL quenching ((3), (4)). Thus, the sample was not annealed. (c) Raman map displaying the spatial distribution of the WS$_2$ A$_{1g}$ mode intensity. Enhancement occurs in regions with pronounced PL quenching. (d) Averaged Raman spectra of heterostructure regions. Average enhancement factors of 3.34~$\pm$~0.53 (3) and 2.26~$\pm$~0.30 (4) are obtained. Maximum detected enhancement factors are 4.73 (3) and 4.26 (4). PL and Raman spectra were measured using an excitation wavelength of 532\,nm.}
	\label{Supplement3}
\end{figure*}


\clearpage
\section{Supplementary theoretical data} 

\begin{figure*} [h]
	\centering
	\includegraphics[width=1.0\textwidth]{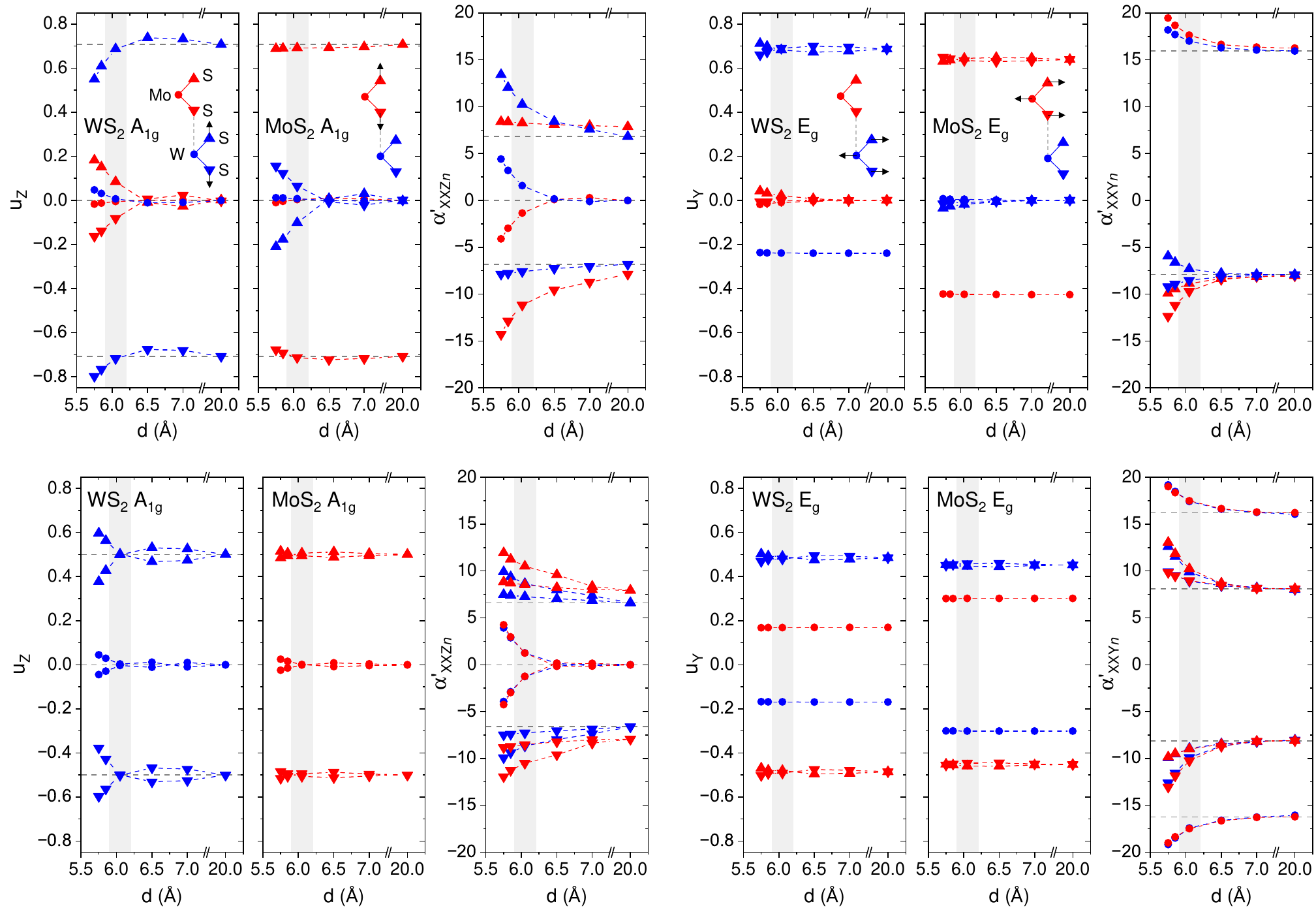}
	\caption{Calculated non-zero contributions to $I_{xx}$ components of Raman intensity tensor for A$_{1g}$ and E$_g$ phonon modes of MoS$_2$ and WS$_2$ layers (top row) in R$^X_h$ MoS$_2$/WS$_2$ heterostructure and (bottom row) in H$^h_h$ MoS$_2$ and WS$_2$ bilayers: $u_z(y)$, the out-of(in)-plane atoms eigendisplacements and $\alpha'_{xxz(y)}$, the atomic-resolved Raman polarizabilities. Blue (red) symbols and lines correspond to WS$_2$ (MoS$_2$) layers: dots (triangles) are used for metal (sulfur) atoms with upper (lower) triangles for upper (lower) sulfur atoms in a layer. Horizontal dashed lines mark some of the separated layers values for reference. Insets present the atomic displacements for the considered phonon modes.} 
	\label{theory-si}
\end{figure*}

\newpage

\clearpage

\end{document}